# Ultra-Thin, Ultra-Light, Rainbow-Free AR Glasses Based on Single-Layer Full-Color SiC Diffrcative Waveguide

Boqu Chen[1,3], Ce Li[5,6], Xiaoxuan Li[2], Ding Zhao[5,6], Lu Cai[4*], Kaikai Du[4,5,6*] and Min Qiu[2,3,5,6*]

[1]. College of Information Science and Electronic Engineering, Zhejiang University, Hangzhou 310007, Zhejiang Province, China

[2]. Key Laboratory of 3D Micro/Nano Fabrication and Characterization of Zhejiang Province, School of Engineering, Westlake University, 18 Shilongshan Road, Hangzhou 310024, Zhejiang Province, China.

[3]. Institute of Advanced Technology, Westlake Institute for Advanced Study, 18 Shilongshan Road, Hangzhou 310024, Zhejiang Province, China.

[4]. Moldnano (Hangzhou) Technology Co., LTD., Hangzhou 311100, Zhejiang Province, China.

[5]. Westlake Institute for Optoelectronics, Fuyang, Hangzhou 311421, Zhejiang Province, China.

[6]. Zhejiang Key Laboratory of 3D Micro/Nano Fabrication and Characterization, Westlake Institute for Optoelectronics, Fuyang, Hangzhou 311421, Zhejiang Province, China.

**Abstract**

As information interaction technology advances, the efficiency, dimensionality, and user experience of information transmission have significantly improved. Communication has evolved from letters to telegraphs, markedly increasing transmission speed; from telephones to video calls, enhancing communication dimensions; and from smartphones to augmented reality (AR) displays, which provide increasingly immersive user experiences. Surface relief grating (SRG) diffractive waveguides have attracted considerable attention for their optimal balance between weight, size, optical performance, and mass production capabilities, positioning them as a leading solution for AR displays. However, as consumer expectations for higher display quality and better device integration rise, traditional high-refractive-index glass-based diffractive waveguides face limitations, including bulkiness, heavy weight, and conspicuous rainbow artifacts in full-color displays. To overcome these challenges, a novel solution: ultra-thin, lightweight silicon carbide (SiC) AR prescription glasses was proposed. This solution achieves full-color displays without rainbow artifacts, with total weight of just 2.685 g and thickness of only 0.55 mm. Moreover, these glasses are compatible with prescription Fresnel lenses and are well-suited for scalable mass production. This innovation provides a robust platform for the seamless integration of augmented reality into daily life, offering significant potential to enhance user interaction.

**Introduction**

With the advancement of technology, the speed, dimensions, and interactivity of information transmission have significantly improved. Alongside the increasing demand for efficient and rapid information exchange, there is a growing expectation for more immersive interactive experiences. In recent decades, AR technology has evolved rapidly, spearheading a revolution in human-computer interaction. AR allows users to experience an interaction between the physical and virtual worlds, where virtual objects are superimposed or integrated into real-world environments[1-3]. Based on different methods of merging virtual images with reality, AR head-mounted displays (HMDs) can be classified into video see-through HMDs[4,5] and optical see-through HMDs.

Optical see-through HMDs are widely adopted due to their portability, real-time capabilities,

and safety advantages. The first optical see-through HMD was proposed by Sutherland in the 1960s[6]. Since then, optical see-through technology has been continuously explored in military[7-11], industrial[12,13], and consumer electronics applications[14-16]. Various approaches have been developed to guide images from micro projector to the observer, integrating real-world views with virtual images[16,17]. Early HMD optical combiners were based on traditional axial beam splitters, as seen in Google Glass[18-20]. However, since the field of view (FOV) and frame size are proportional to the size of the optical elements, achieving a balance between performance and comfort led to smaller FOVs in such smart glasses. To achieve a larger FOV, HMDs using off-axis aspheric mirrors were introduced[21], but the large size and off-axis aberrations compromised user comfort. Later, full internal reflection (TIR) solutions based on freeform optics[22] were proposed, offering high-quality displays with a large FOV (e.g., Canon[23-25]). However, the bulky form factor limited their adoption. To achieve a satisfactory display with an acceptable FOV and compact size, waveguide HMDs with cascaded mirror arrays[26-28] and light-guiding optical elements (such as Lumus[29-31]) were developed. This system, based on reflection, results in relatively low optical aberrations, but the high manufacturing precision and coating quality required remain difficult problem. Compared to traditional bulky refractive optics, planar diffractive waveguides have become a recent research focus due to their excellent trade-offs between form factor, FOV, frame size, and mass production[32,33]. The optical system works by coupling light from a micro projector into a thin plate via a small coupling area[34], then projecting the image to the human eye through pupil expansion and outcoupling area[35]. With the continuous advancements in optical design and micro-nano fabrication, diffractive waveguides are increasingly applied in military[36,37] and consumer electronics, gradually becoming the optimal choice for future development.

As shown in Figure 1, a mass-producible, ultra-light, and thin SiC AR prescription glasses with 30° FOV, full-color, rainbow-artifacts-free display was proposed. SiC was chosen for its high refractive index, excellent thermal conductivity, low optical loss, low density, mechanical hardness, and chemical resistance across the visible to near-infrared spectrum[38-41]. To address the limitations of traditional nanoimprinting technologies, a mass-production-compatible nanoimprint lift-off process was developed, enabling the large-scale transfer of metal patterns onto SiC wafers. The SiC waveguide was fabricated by transferring the patterns from the metal mask to the SiC surface using dry etching. To enhance transmittance and protect the waveguide structure, a novel ultra-thin packaging process was introduced, employing a sandwich structure of hard coating and anti-reflective coatings to enclose the SiC waveguide. After packaging and laser cutting, the final monolithic SiC waveguide weighs only 2.7 g with a thickness of 0.55 mm, representing a significant improvement over current mainstream AR glasses. Experimental results demonstrated that the SiC waveguide achieves full-color, rainbow-free display at a 30° FOV, with potential for further FOV expansion as optical system performance improves. To meet the refractive correction needs of myopic users, ultra-thin vision correction lenses were developed, which can be attached to the SiC waveguide surface, providing low-cost, high-performance visual correction functionality. The SiC AR glasses achieve full-color display without requiring multiple waveguides, addressing the industry's challenge of rainbow artifacts while optimizing weight, volume, and vision correction. This innovation advances diffractive waveguide display technology, providing a clear direction for the future of information interaction.

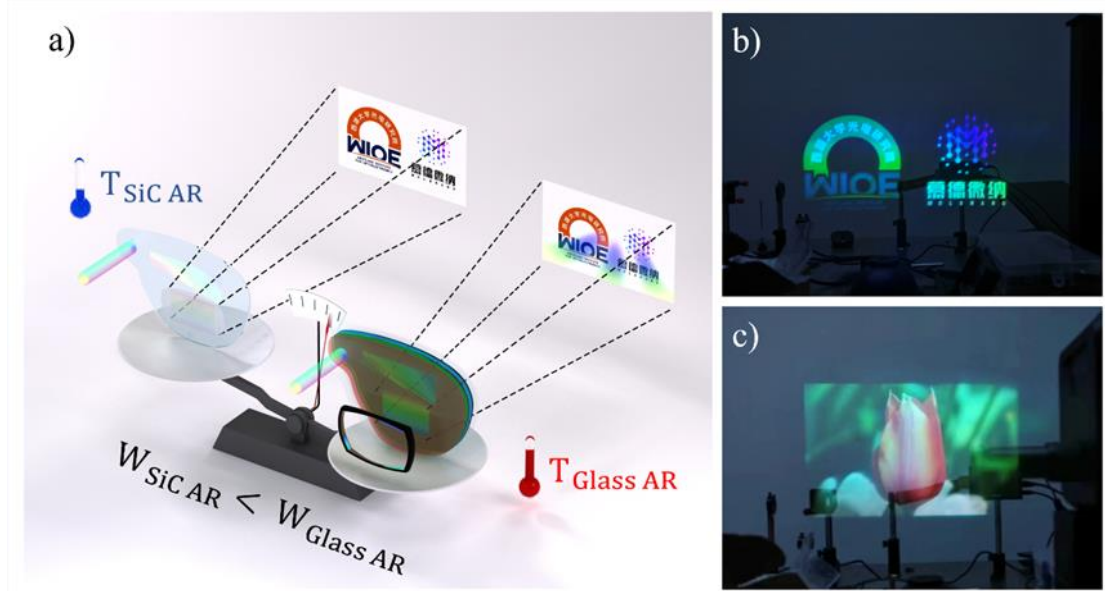

**Figure 1: Operational schematic of SiC AR glasses. (a)** Comparison of SiC AR glasses (left) and conventional AR glasses (right) in achieving full-color display and vision correction. SiC AR glasses demonstrate advantages in being lighter, thinner, offering superior heat dissipation, and eliminating rainbow artifacts. **(b-c)** Photographs of the full-color display produced by SiC AR glasses.

## Design and Principle

The proposed structure for an ultra-thin colored light waveguide in AR optical combiners is a hard coatings, SiC waveguide, and anti-reflect coating sandwich configuration, referred to simply as the SiC waveguide. This design employs a SiC layer as the waveguide to fulfill the requirements for light turning and exit pupil expansion (EPE) for RGB light. Hard coating layers are applied on both sides of the SiC waveguide, replacing the traditional cover layer to protect the waveguide. Meanwile, hard coating can eliminates the gap in conventional AR waveguide and reduces weight. Figure 2a-b shows the overall layout of the waveguide and the corresponding k-space, respectively. Figure 2b illustrates the k-space image of the SiC waveguide: the inner circle representing the k-vector in an air environment, and the dashed middle circle for the k-vector of the hard coatings layer, and the outer RGB circles corresponding to the k-vectors of red, green, and blue light, respectively. All k-vectors are normalized by $(2\pi/\lambda)$. The central rectangle indicating the k-vector range of the FOV. The grating vectors for the IC, EPE, and OC regions form angles of -60°, 60°, and 0° with the horizontal axis, respectively. Light from the projection system, confined within the k-vectors of the FOV, enters the IC region of the SiC waveguide. It is first modulated by the IC region grating, then undergoes TIR within the SiC layer to reach the EPE region. Continuing through the EPE, the light reaches the OC, where it is modulated by the OC grating and coupled out of the SiC waveguide to be captured by the human eye. The k-space image demonstrates that RGB light within the FOV range can undergo TIR in the SiC layer and be coupled out in the same direction as the incident light, this confirms the theoretical feasibility of using the SiC waveguide for AR applications.

In addition, due to the smaller period design of SiC waveguide, rainbow artifacts can be suppressed effectively. Rainbow artifacts are an optical phenomenon in waveguides where ambient light diffracted by the grating on the waveguide surface disperses, creating bright regions that resemble rainbows and significantly degrade the user experience of AR glasses. Reducing or suppressing rainbow has become one of the most important issues for AR glasses to achieve widespread market adoption. In SiC waveguide, ambient light can be diffracted at a larger angle, directing it mostly outside the human eye's line of sight, which effectively suppresses rainbow artifacts. Figures 2c-d illustrate the rainbow suppression effect in the SiC waveguide, where the large rectangle represents the k-vector angles that the eye can perceive within the Eye-box. The blue ring denotes that the ambient light, after being diffracted by the grating in OC region, produces outgoing light rays that are positioned outside the angle receivable by the Eye-box, ensuring that rainbow artifacts are not visible to the human eye. For a clearer comparison, consider a waveguide with a refractive index of n=1.9. To achieve a single-pane full-color display with a 30° FOV, the grating period is calculated as P=370 nm. Under the same Eye-box conditions, some blue and green components of ambient light diffracted by this grating enter the eye's reception area, indicating a potential for rainbow artifacts to be observed in such a waveguide structure.

Using independently developed code for waveguide structure optimization, the coupling efficiency for RGB colors across various FOV angles can be calculated, as shown in Figures 2e-g. Binary grating structure with 262 nm period are selected for all grating regions. The optimized grating structure parameters are: for the IC region, duty cycle is 58%, grating height is 169 nm; for the EPE region, duty cycle 50.5%, grating height is 167 nm; for the OC region, duty cycle 42.1%, grating height 27 nm. Figures 2e-g demonstrate that, with these parameters, the coupling efficiency for all angles and colors is greater than zero, allowing for the output of a complete full-color image across the entire FOV. Furthermore, to enhance the uniformity of the output image, further partitioning of the EPE and OC regions could be implemented to improve image quality.

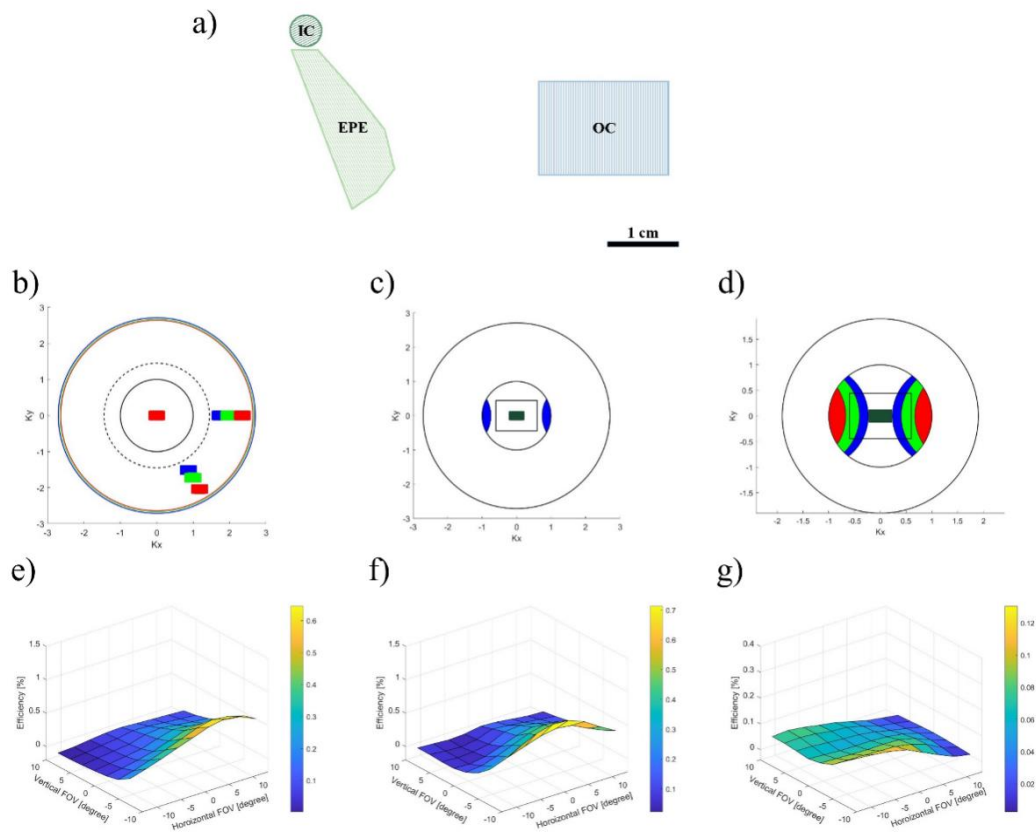

**Figure 2: Optical simulation of SiC AR glasses. (a)** Layout distribution of SiC AR. **(b)** The k-space and k-vectors of SiC AR. **(c)** Diffraction of ambient light in SiC AR k-space does not overlap with the eye-box, eliminating the rainbow artifacts. **(d)** Overlap

of ambient light diffraction with the eye-box in conventional AR (n=1.9), leading to the rainbow artifacts. **(e-g)** Outcoupling efficiency of SiC AR at 456, 520, and 617 nm wavelengths within the FOV range.

## Results and Discussion

To evaluate the breakthrough performance of SiC AR glasses, a series of static and optical tests were designed, as shown in Figure 3. The large-area homogeneous SiC waveguides were fabricated using the custom-developed NIL-to-lift-off process on 4-inch 4H-SiC wafers (Figure 3a). This process effectively addressed challenges associated with the conventional NIL process, specifically the low etch selectivity of nanoimprint resists and contamination from residuals in the NIL-to-etch method. The process enabled the precise replication of homogeneous SiC nano-gratings with a period of 262 nm, covering a large structural area of 3.34 cm². SEM images (Figure 3b) confirm that the nano-grating structure maintained excellent morphology and dimensional accuracy, matching the design specifications with high precision. These results demonstrate the feasibility of mass production for dense, small-period micro-nanostructures in hard-to-etch materials, such as SiC, introducing a scalable method for fabricating high-hardness materials in industrial applications.

In addition to waveguide fabrication, the focus was on protecting the SiC waveguide and improving light transmission by developing an ultra-thin packaging method. This approach used a sandwich structure composed of a hard coating and anti-reflective coating, applied through spin-coating and deposition, respectively. The SiC AR glasses were laser-cut into lens shapes (Figure 3c) to allow for evaluation in practical use cases. Weight measurements using a high-precision balance showed that a single SiC AR lens weighs only 2.6856 g (Figure 3d), significantly lighter than conventional AR glasses, which typically weigh 10–15g per lens. Additionally, the thickness of the SiC AR glasses was measured to be just 0.55 mm using a vernier caliper and optical microscope (Figure 3e), much thinner than the 2–3 mm thickness of mainstream AR glasses made with high-refractive-index glass or resin. These reductions in size and weight highlight the SiC AR glasses' potential for enhanced portability, improved comfort, and easier integration into compact AR systems.

To evaluate the optical performance of the SiC AR glasses, a custom optical testing platform was designed and assembled. This setup, mounted on a vibration isolation table, included a DLP light engine, a luminance colorimeter mounted on a six-axis translation stage, and a sample holder for the AR glasses (Figure 3f). The light engine projected images through the optical path to the sample holder, where the SiC AR glasses were positioned. By aligning the AR glasses with the projector's optical path, proper waveguide coupling was ensured. The luminance colorimeter was adjusted to capture the light exiting the waveguide, allowing for measurements at nine points within the display FOV.

The results of the optical tests demonstrated excellent brightness uniformity across the display. Specifically, the brightness uniformity for RGB mixed light was measured at 3.18%, with red at 4.68%, green at 2.82%, and blue at 3.57%. The luminous efficiency for RGB mixed light was measured at 175.45 nit/lm, with red at 163.36 nit/lm, green at 179.13 nit/lm, and blue at 148.95 nit/lm. Despite the absence of specific optimizations for display uniformity, the measured MTF@22PPD score was 0.2669 (horizontal) and 0.3055 (vertical), indicating that the waveguide was capable of achieving high-quality full-color displays. Further improvements in uniformity could be achieved by implementing structural duty cycle or height modulation. The FOV for the SiC AR glasses was measured at 45.7°, slightly higher than the design value due to the 50° FOV of the projector. Additionally, distortion was measured at 1.48%, well within the acceptable range and consistent with design targets. The comprehensive optical performance tests confirmed that the SiC AR glasses can deliver full-color displays with high FOV and minimal distortion. These features, combined with the device's lightweight and thin form factor, offer significant advantages over conventional AR glasses. The SiC AR technology provides a promising solution for next-generation diffractive waveguide displays, offering improved optical performance, enhanced portability, and the potential for mass production in commercial applications.

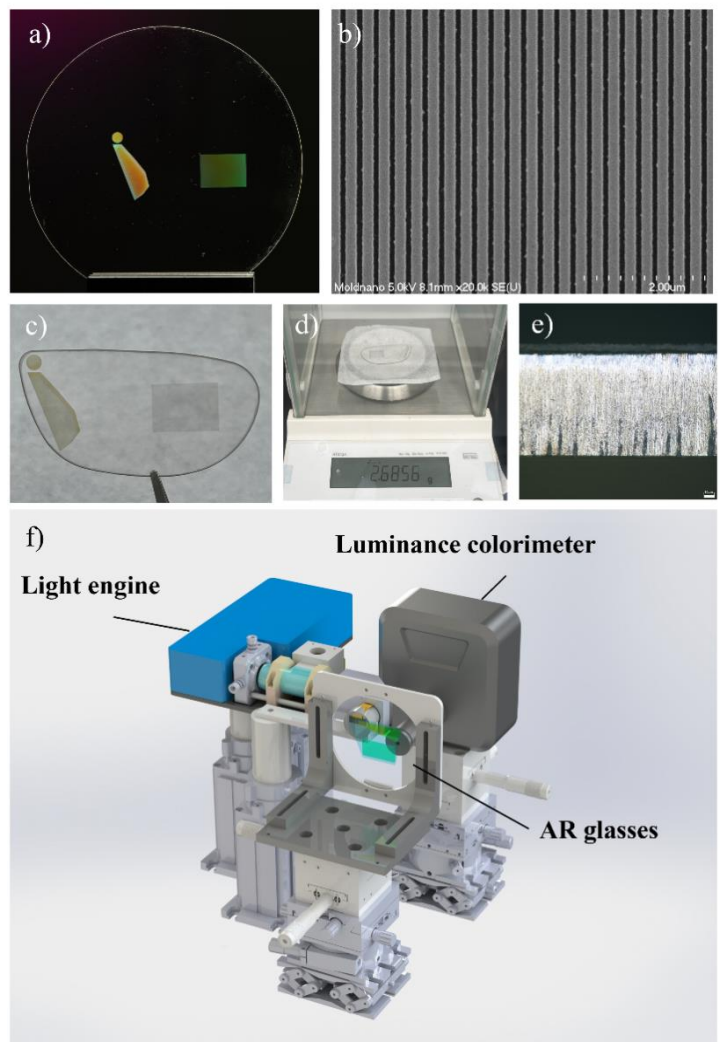

**Figure3. Static and optical test of SiC AR glasses. (a)** SiC AR waveguide. **(b)** SEM diagram of SiC nano-grating. **(c)** SiC AR glasses after packaging and laser cutting. **(d)** Weight measurement using a high-precision balance. **(e)** Cross-sectional image of the SiC AR glasses. **(f)** Optical testing setup, including the light engine, luminance colorimeter, and multi-axis sample holder.

## Materials and methods

The SiC AR glasses consist of a SiC waveguide and a Fresnel prescription lens, with the fabrication process illustrated in Figure 4. Conventional diffractive waveguide AR glasses use NIL for mass production, but the soft imprint resist materials are prone to damage. To enhance display performance and durability, an imprint-to-etch process transfers the pattern from the resist to the substrate. However, this method is limited by rapid resist consumption during etching, which constrains structure depth. This limitation was addressed by developing a NIL to lift-off process, enabling mass production of metal masks that prevent defects from residual resist during etching and improve cleanliness. In addition, bulky resin covers were replaced with an ultra-thin sandwich cover featuring anti-reflective and hard coatings, which protect the nano-gratings while improving light transmission. For vision correction, an ultra-thin Fresnel lens was developed using diamond turning to create NIL masters, allowing for mass production of prescription lenses that integrate seamlessly with the waveguide cover. These optimized processes enabled the successful integration of SiC waveguides, ultra-thin packaging, and planar prescription lenses.

The NIL master was created using electron beam lithography (EBL) and inductively coupled plasma (ICP) etching to form high-precision grating structures on a silicon substrate (Figure 4, **NIL mask fabrication**). A 100 kV EBL system (Raith EBPG) was used to pattern a 4-inch silicon wafer. A high-current 10 nA exposure improved writing speed while maintaining morphology and accuracy. After development, the EBL resist (ZEP 520A) acted as an etch mask, allowing the target depth to be etched into the substrate. Residual resist was removed via NMP flushing and oxygen plasma cleaning, completing the NIL master fabrication.

To extend the lifespan of the NIL master and improve release properties, the pattern was transferred to a PET film, which served as the working stamp for subsequent imprinting (Figure 4, **Working stamp fabrication**). After anti-adhesion treatment, the master was coated with working stamp resist, UV-cured, and pressed onto the PET film. The PET film was then separated, retaining the inverse pattern for future imprinting.

For SiC waveguide mass production, a NIL to lift-off process was developed (Figure 4, **Waveguide mass production**), addressing the limitations of conventional NIL in fragility and optical performance. Unlike the NIL-to-etch process, this method utilizes a metal mask, providing higher etch selectivity and broader applicability. The process begins with spin-coating lift-off resist on a SiC wafer, followed by $SiO_2$ deposition and NIL resist application. After imprinting and UV curing, the NIL residual layers, $SiO_2$, and lift-off resist were removed through multiple etching steps. To ensure high lift-off success, reactive ion beam etching (RIBE) was used to remove residual materials, leaving a clean photomask. Chromium was deposited onto the desired area, and non-patterned areas were lifted off. The nano-grating structure was then etched into the SiC substrate using ICP, and the chromium mask was subsequently removed.

To improve optical transmission and protect the waveguide, a novel ultra-thin packaging process was introduced, involving a sandwich structure with anti-reflective and hard coatings (Figure 4, **SiC AR packaging process**). A 30 μm protective layer was formed by spin-coating hardening liquid onto the waveguide, followed by UV curing. Alternating layers of $SiO_2$ and $TiO_2$ were deposited to form a multi-layer anti-reflective coating.

With the growing demand for vision correction in AR devices, prescription lenses are becoming critical. Current solutions add custom lenses in front of the waveguide, increasing weight and cost. This issue was addressed by using diamond turning to fabricate Fresnel lens masters, enabling low-cost mass production through NIL. Ultra-thin Fresnel lenses were affixed to the lens surface, providing efficient vision correction with minimal size and cost (Figure 4, **Prescription Fresnel lens fabrication**).

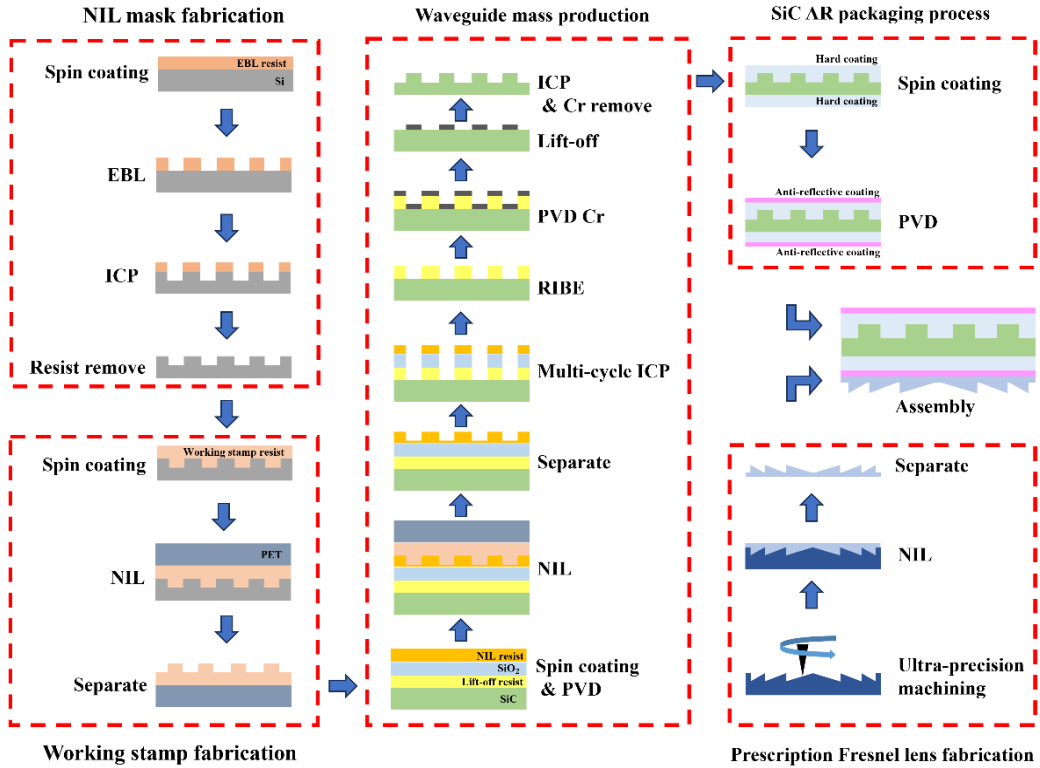

Figure4. SiC AR fabrication flow.

## Conclusion

In conclusion, the experiments demonstrated that SiC AR glasses provide lightweight, thin, full-color display capabilities without rainbow artifacts. The design objectives were successfully met, achieving full-color display via a monolithic waveguide while effectively eliminating rainbow artifacts. To enable scalable fabrication of homogeneous nano gratings on SiC substrates, a NIL to lift-off process was developed, allowing for the mass production of waveguides on 4-inch 4H-SiC wafers. To safeguard the SiC waveguide and improve optical transmission, an ultra-thin packaging technique was implemented, utilizing spin-coated hard coating and anti-reflective coatings, achieving encapsulation with a thickness of just 30 μm per side. Following laser cutting, each SiC AR lens weighed 2.6856 g and measured 0.55 mm in thickness—representing a significant enhancement over conventional full-color AR glasses in terms of both size and weight. In optical tests, the SiC AR glasses demonstrated a wide field of view (FOV) with minimal image distortion, while maintaining full-color display quality. These findings underscore the superior performance of SiC AR glasses compared to conventional AR systems, which often rely on more complex designs and higher fabrication complexity to achieve similar results.

The techniques developed in this work open up possibilities for high-performance devices compatible with semiconductor manufacturing, particularly through the use of materials with exceptional optical and mechanical properties, such as diamond. While the current SiC AR design already achieves monolithic full-color display, further enhancements in display brightness and color uniformity could be realized by modulating the structural duty cycle and depth in specific regions. Additionally, adopting a two-dimensional pupil expansion strategy could lead to a more compact design while enabling an even larger FOV. By integrating these optimization approaches, future optical devices with enhanced performance could be developed for a range of applications, including augmented reality, metasurfaces, and aerospace. This advancement addresses critical optical challenges in modern information interaction and industrial systems, positioning SiC AR technology as a key player in next-generation optical solutions.